# A Deep Reinforcement Learning-Based Controller for Magnetorheological-Damped Vehicle Suspension

AmirReza BabaAhmadi[1*] , Masoud ShariatPanahi[2] , Moosa Ayati[3]

[1,2,3] School of Mechanical Engineering, College of Engineering, University of Tehran, Tehran, Iran

[1]Babaahmadi.amir@ut.ac.ir , [2]mshariatp@ut.ac.ir , [3]m.ayati@ut.ac.ir

## Abstract

This article presents a novel approach to designing a controller for a vehicle suspension system equipped with magnetorheological (MR) dampers. The proposed method centers around the concept that an optimal control strategy can be acquired by employing a reinforcement learning algorithm with continuous states and actions, utilizing either real-world or simulated experiments. The Twin Delayed Deep Deterministic Policy Gradient (TD3) algorithm is employed to process the sensor data and generate the appropriate actuation voltage for the MR damper. To evaluate the system's performance, a quarter vehicle model is utilized, incorporating the modified Bouc-Wen MR damper model. This model facilitates the calculation of important suspension metrics such as displacement, sprung mass acceleration, and dynamic tire load within the suspension working space. The reward function utilized in the deep reinforcement learning algorithm is based on the sprung mass acceleration. Through numerous simulated experiments, the results indicate that the proposed approach surpasses conventional suspension control strategies in terms of both ride comfort and stability.

## 1. Introduction

As one of the most critical components of the vehicle, the suspension system significantly improves ride comfort and road holding, preventing damage and reducing passenger fatigue. Suspension systems are classified as passive, active, or semi-active. Due to the fixed and unchanging nature of passive suspension parameters, passive suspension cannot guarantee ride comfort and stability when the environment or suspension parameters vary. Thus, active and semi-active suspension systems with tunable parameters can compensate for the limitations mentioned above of passive suspension systems. The uncertainty inherent in the suspension system's road roughness and parameter variation in real applications is unavoidable.

Magnetorheological (MR) fluid dampers are adaptive controllable devices that have attracted a considerable amount of interest due to their attractive features, such as simplicity, low power consumption, and high force capacity. Their broad applicability is evident in various industries, including the automotive industry [1][2], civil structures [3][4], and railway vehicles [5]. MR dampers are classified as semi-active devices because they can produce damping force by simply applying voltage to their coils without the need for a mechanical mechanism. As a result, two distinct controller types are required to regulate semi-active suspensions. First, the system controller calculates the required damping force to ensure ride comfort and road holding simultaneously. The system controller inputs are derived from the suspension system's state feedback. Second, the damper controller determines the voltage that should be applied to the MR-damper for its current force to track the force specified by the system controller.

Several control techniques have been developed for semi-active suspension systems, aiming to enhance ride comfort and stability. These techniques include $H\infty$ control [6], [7], [8], skyhook control [9], [10], [11], adaptive control utilizing neural networks [12], robust control [13], LQG control [14], [15], and optimal PID control [16][17]. While PID controllers are known for their simple structure, they prove to be ineffective in semi-active suspension systems due to uncertain parameters. One well-known semi-active suspension control method is the skyhook control, which is described in [18]. This method stands out for its simplicity in terms of structure, ease of implementation, and acceptable performance. However, the presence of time delay significantly impacts suspension performance, often leading to instability and wheel jump. In an attempt to address this issue, the authors of developed an optimal LQG controller. Nevertheless, the control parameters were calculated without taking into account the uncertain factors in the system modeling.

When the system's parameters change sufficiently, the system becomes unstable. The authors of [19] and [20] aimed to address the issue of the blind design of the fuzzy controller, where this was, in fact, a PID controller, and a PID was designed using rule description. Developing an effective fuzzy control system requires a thorough understanding of the underlying system. The sliding mode controller presented in [21] demonstrates commendable performance and robust behavior. However, one critical drawback of sliding mode control is the occurrence of chattering, which can lead to instability due to the activation of high-frequency modes in the controlled system. While the adaptive controller discussed in [22] enhances ride comfort, while its performance in terms of road holding is unsatisfactory. In an alternative approach, reference [23] introduced a fast model prediction controller (FMPC), while the authors of [24] proposed a robust model prediction controller (RMPC).

Both of the aforementioned predictive controllers utilized road models in advance of designing controllers. The primary goal of developing a predictive controller is to provide highly accurate information from the system, which is impossible when driving on various roads with high uncertainty. In [21], from another perspective, it is demonstrated that neural-networked-based controllers can solve complex and nonlinear problems. However, like many other supervised learning algorithms, neural networks require a large number of labeled samples. When developing the control effort, only the current state is considered; future states are not considered. These critical issues impose constraints on the use of neural network controllers. Reference, [22] which we discssued before proposes a novel nonlinear adaptive smart controller based on the classification of road profiles. The primary disadvantage of this method is that a new classifier must be constructed when the control strategy changes.

In recent years, several research papers have aimed to tackle the aforementioned issues. In [25], a novel robust H∞/H2 controller is proposed, taking into account the system's delay response. Another study [26] introduces a new Fuzzy-PID controller specifically designed for real-time control of semi-active suspension systems equipped with MR dampers. In a recent investigation [27], the authors developed an adaptive sliding mode fault-tolerant controller for the semi-active suspension system, capable of addressing uncertainties within the system. Additionally, they employed a fuzzy model and conducted multiple simulations to validate their findings.

With the advent of Deep Learning (DL) techniques that overcome several of the significant limitations of classical machine learning algorithms, some researchers attempted to apply DL algorithms to the suspension control problem. The authors of [28] propose a DDPG-based algorithm for a particular type of suspension system. The primary drawback of this algorithm is its inability to locate the globally optimal solution to a problem.

Two types of controllers are required to control a semi-active suspension system. A system controller analyzes the system's feedback samples and predicts the desired damping force. The damper controller receives the system controller's predicted force and suspension system displacement; it then predicts the applied voltage exerted on damper coils.

The paper proposes an improved DRL controller that combines a system controller and a damper controller to provide the best ride comfort and stability possible. Notably, MR-based suspension systems operate in two modes: open-loop mode (without the use of a controller) operates with a constant damping coefficient, similar to passive suspension systems. Both the system and damper controllers work in a closed-loop system to adjust the damping force in response to road conditions by tuning the required damper's voltage input.

## 2. MR-damped suspension model

A simplified quarter-vehicle model with a semi-active suspension is shown in Fig. 1, where $m_b$ represents the vehicle's body mass, $m_w$ represents the wheel mass, and $x_b$ and $x_w$ denote body displacement and wheel displacement, respectively. The road profile is denoted by $x_r$. The spring stiffness of the suspension system is $k_s$, and the tire spring stiffness is denoted by $k_t$. We exclude tire damping from this study due to its negligible value. Table 1 contains parameters taken from [29]. Newton's second law is applied to the quarter-model of the vehicle to derive the following equations:

$$m_b \ddot{X}_b + K_s(X_b - X_w) + f = 0 \tag{1}$$

$$m_w \ddot{X}_w - K_s(X_b - X_w) + K_t(X_w - X_r) - f = 0 \tag{2}$$

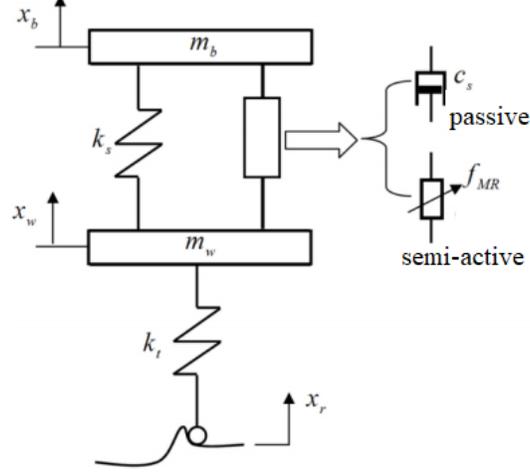

**Fig. 1.** A quarter-car model with two operational modes for a semi-active suspension system (passive and semi-active)

Where the following equation gives the damping force generated by the MR device:

$$f = \begin{cases} C_s(\dot{X}_b - \dot{X}_w) & \text{for passive suspension} \\ f\,MR & \text{for semi} - \text{active suspension} \end{cases} \tag{3}$$

$C_s$ is a coefficient describing the MR passive suspension mode of operation in an MR-damper.

The state-space formulation of the semi-active suspension system is established as follows [17]:

$$\dot{w} = Aw + Bf_{MR} + Dx_r \tag{4}$$

$$w = [x_b \quad x_w \quad \dot{x}_b \quad \dot{x}_w]^T \tag{5}$$

$$A = \begin{bmatrix} 0 & 0 & 1 & 0 \\ 0 & 0 & 0 & 1 \\ -\frac{K_s}{m_s} & \frac{K_s}{m_s} & 0 & 0 \\ \frac{K_s}{m_w} & -\frac{K_s+K_t}{m_w} & 0 & 0 \end{bmatrix} \tag{6}$$

$$B = \begin{bmatrix} 0 & 0 & -\frac{1}{m_s} & \frac{1}{m_w} \end{bmatrix}^T \tag{7}$$

$$D = \begin{bmatrix} 0 & 0 & 0 & \frac{K_t}{m_w} \end{bmatrix}^T \tag{8}$$

**Table 1.** Semi-active suspension parameters [29]

| Parameter | Symbol | Value (Unit) |
|---|---|---|
| Vehicle body mass | $m_b$ | 375(kg) |
| Vehicle wheel mass | $m_w$ | 29.5 (kg) |
| Suspension stiffness | $k_s$ | 20.58 (KN/m) |
| Damping coefficient | $C_s$ | 772 (Ns/m) |
| Tire stiffness | $k_t$ | 200 (KN/m) |

The force generated by a magnetorheological (MR) damper is represented by $f_{mr}$ and is dependent on the time-varying external voltage applied to its magnetic coil, denoted by $V$, as well as the relative displacement between the sprung mass and unsprung mass, which is referred to as suspension working space (SWS) and denoted by $x = x_b - x_w$. The calculation of $f_{mr}$ is based on the modified Bouc-Wen model, as presented in Eq. 9, which has been derived from previous research [30] and incorporated into the suspension system illustrated in Fig. 1.

$$F(t) = c_1 \dot{y} + k_1 (x - x_0) \tag{9}$$

$$\dot{z} = -\gamma|\dot{x} - \dot{y}||z||z|^{n-1} - \beta(\dot{x} - \dot{y})|z|^n + A(\dot{x} - \dot{y}) \quad (10)$$

$$\dot{y} = \frac{1}{c_1 + c_0}\{\alpha z + k_0(x - y) + c_0\dot{x}\} \quad (11)$$

$$\alpha = a_a + a_b u \quad (12)$$

$$c_1 = c_{1a} + c_{1b} u \quad (13)$$

$$c_0 = c_{0a} + c_{0b} u \quad (14)$$

$$\dot{u} = -\eta(u - v) \quad (15)$$

The internal movement of the MR-damper is denoted as $y$. $u$ is the output signal from a first-order filter, and $z$ is a parameter that guarantees the hysteretic behavior of MR fluid. Accumulator stiffness is denoted by $k_t$. $C_0$ and $C_1$ represent viscous damping at high and low velocity for the MR-damper, respectively. The stiffness regulator at high damper velocity is denoted by $K_0$. MR damper accumulator simulation is performed via $X_0$. The scale and shape of the hysteresis behavior of the MR-damper parameters for the simulation are $\gamma, \beta, \delta$ and $\eta$, respectively.

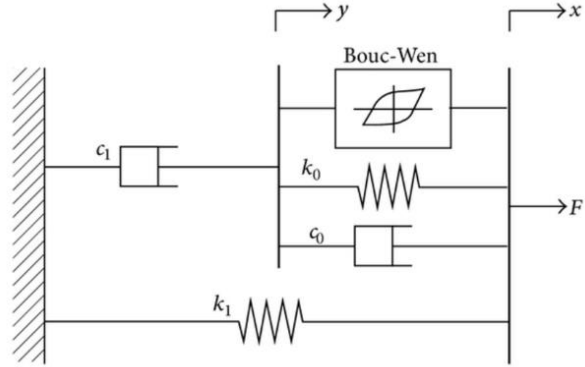

**Fig. 2.** Modified Bouc-Wen model for the MR-damper

Table 2 contains the parameters for the modified Bouc-Wen MR-damper model, as adapted from a previous study [30].

Table 2. Modified Bouc-Wen parameters for the MR-damper [30]

| Parameter | Value | Parameter | Value |
|---|---|---|---|
| $c_{0a}$ | 784 $Nsm^{(-1)}$ | $\alpha_b$ | 38430 $NsV^{(-1)}m^{(-1)}$ |
| $c_{0b}$ | 1803 $NsV^{(-1)}m^{(-1)}$ | $\beta$ | 2059020 $m^{-2}$ |
| $c_{1a}$ | 14649 $Nsm^{(-1)}$ | $\gamma$ | 136320 $m^{-2}$ |
| $c_{1b}$ | 34622 $NsV^{(-1)}m^{(-1)}$ | $\eta$ | 190 $s^{-1}$ |
| $k_1$ | 840 $Nm^{(-1)}$ | A | 58 |
| $k_0$ | 3610 $Nm^{(-1)}$ | n | 2 |
| $\alpha_a$ | 12441 $Nm^{(-1)}$ | $x_0$ | 0.245 |

## 2.1. Control Strategy based on DRL-TD3

As previously stated, the Twin Delayed Deep Deterministic Policy Gradient (TD3) was used as the universal controller for the MR-damped suspension system. TD3 is identical to DDPG but incorporates three additional features to address DDPG issues.

- *Clipped double Q learning*: In TD3, we compute the Q value using two critic networks and the target value using two target networks. We used only one critic and one target network in DDPG. In TD3, we use two target critic networks to compute two target Q values. Then, when computing the loss function, we choose the smaller of the two. This will ensure that we do not overestimate the target Q value.
- *Delayed policy update*: Unlike DDPG, we added a delay to the actor-network parameter update in TD3. While the parameters of the critic networks are updated at each step of the episode, the actor-network (policy network) is delayed and updated after every two steps.
- *Target policy smoothing*: In DDPG, the algorithm generates distinct target values for identical actions. As a result, the target's variance would be high. Thus, we reduce variance by adding noise to the target action.

This section provides additional information about Twin Delayed Deep Deterministic Policy Gradients (TD3). In TD3, we employ six artificial neural networks, four of which are critic networks and two of which are two-actor networks.

- The main critic neural network parameters are denoted by $\theta 1$ and $\theta 2$
- The two target critic neural network parameters are represented by $\theta 1'$ and $\theta 2'$
- The main actor-network parameter is denoted by $\phi$
- The target actor-network parameter is denoted by $\phi'$

Initially, we must initialize the two main critic network parameters, $\theta 1$ and $\theta 2$, and the main critic network parameter $\phi$ with random values. Since the target network parameter is only a copy of the main network parameter, the two target critic network parameter $\theta 1'$ and $\theta 2'$ by copying $\theta 1$ and $\theta 2$, are simply initialized, respectively. Similarly, we initialize the target actor parameter $\phi'$, by just copying the main actor-network parameter $\phi$. Also, the replay buffer is initialized too. Now, we select action $a$ using the actor-network: $a = \mu_\phi(s)$. Rather than selecting the action directly, some noise $\in$ is added to ensure exploration when $\in \sim N(0,\sigma)$. Accordingly, the output action is expressed as follows: $a = \mu_\phi(s) + \in$ (16).

Then, after performing action $a$, we move to the next state $s'$ and obtain reward $r$. This transition information is stored in the replay buffer. During the next step, we randomly sample a minibatch of $K$ transition $(s, a, r, s')$ from the replay buffer. The $K$ transition will be used for updating both the critic and actor network.

The loss function of the critic network is expressed as follows:

$$J(\theta_j) = \frac{1}{k} \Sigma (y_i - Q_{\theta_j}(s_i, a_i))^2 \quad for \ j = 1, 2 \tag{17}$$

In the preceding equation, the following applies:

The action $a_i$ is the action produced by the actor-network as follows:

$$a_i = \mu_\phi(s) \tag{18}$$

$y_i$ is the target value of the critic, that is

$$y_i = r_i + \gamma \min_{j=1,2} Q_{\theta_j'}(s', \tilde{a}) \tag{19}$$

, and the action $\tilde{a}$ is the action produced by the target critic network:

$$\tilde{a} = \mu_{\phi'}(s_i') + \in \ where \in \sim (N(0,\sigma), -c, +c) \tag{20}$$

After computing the loss of the critic network, the gradients $\nabla_{\theta_j} J(\theta_j)$ is computed, and then the critic network parameters will be updated using the gradient descent method.

$$\theta_j = \theta_j - \alpha \nabla_{\theta_j} J(\theta_j) \quad for \ j = 1, 2 \tag{21}$$

Now, the actor network must be updated. The objective function of the actor-network is as follows:

$$J(\phi) = \frac{1}{k}\sum_i Q_{\theta_i}(s_i, a) \tag{22}$$

It must be noted that in Eq. 22, we only use state ($s_i$) from the sampled $K$ transitions $(s, a, r, s')$. The action $a$ is selected by the actor-network.

$$a_i = \mu_\phi(s) \tag{23}$$

In order to maximize the objective function, the gradient of the objective function $\nabla_\phi J(\phi)$ is computed, and the parameters of the network are updated using the gradient ascent:

$$\phi = \phi + \alpha \nabla_\phi J(\phi) \tag{24}$$

We delay the update rather than updating the actor-parameters networks at each time step of the episode. If the time step of the episode is denoted by $t$ and $d$ denotes the number of time steps which we want to delay the update, it can be described as follows:

1- If $t$ mod $d=0$, then:
   1. Compute the gradient of the objective function $\nabla_\phi J(\phi)$
   2. Update the actor-network parameter using the gradient ascent method (24).

Finally, the parameters of the target critic network $\theta 1'$ and $\theta 2'$ and the parameters of the target actor-network $\phi'$ will be updated via soft replacement:

$$\begin{cases} \theta'_j = \tau\theta_j + (1-\tau)\theta'_j \; for \; j = 1,2 \\ \phi' = \tau\phi_j + (1-\tau)\phi' \end{cases} \tag{25}$$

There is a small change in updating the parameters of the target networks. As with the actor-network parameter, we delay updating it for $d$ steps; similarly, we update the target network parameters for every $d$ step; in this case, we can say:

1- If $t$ mod $d=0$, then:
   1. Compute the gradient of the objective function $\nabla_\phi J(\phi)$ and update the actor-network parameter using gradient ascent.

$$\phi = \phi + \alpha \nabla_\phi J(\phi) \tag{26}$$

   2. Update the target critic network parameter and target actor-network parameter as

$$\theta'_j = \tau\theta_j + (1-\tau)\theta'_j \text{ for } j = 1,2 \tag{27}$$

and

$$\phi' = \tau\phi + (1-\tau)\phi' \tag{28}$$

, respectively.

The previous steps for several episodes must be repeated to improve the policy. The following pseudocode is prepared to understand better how TD3 works:

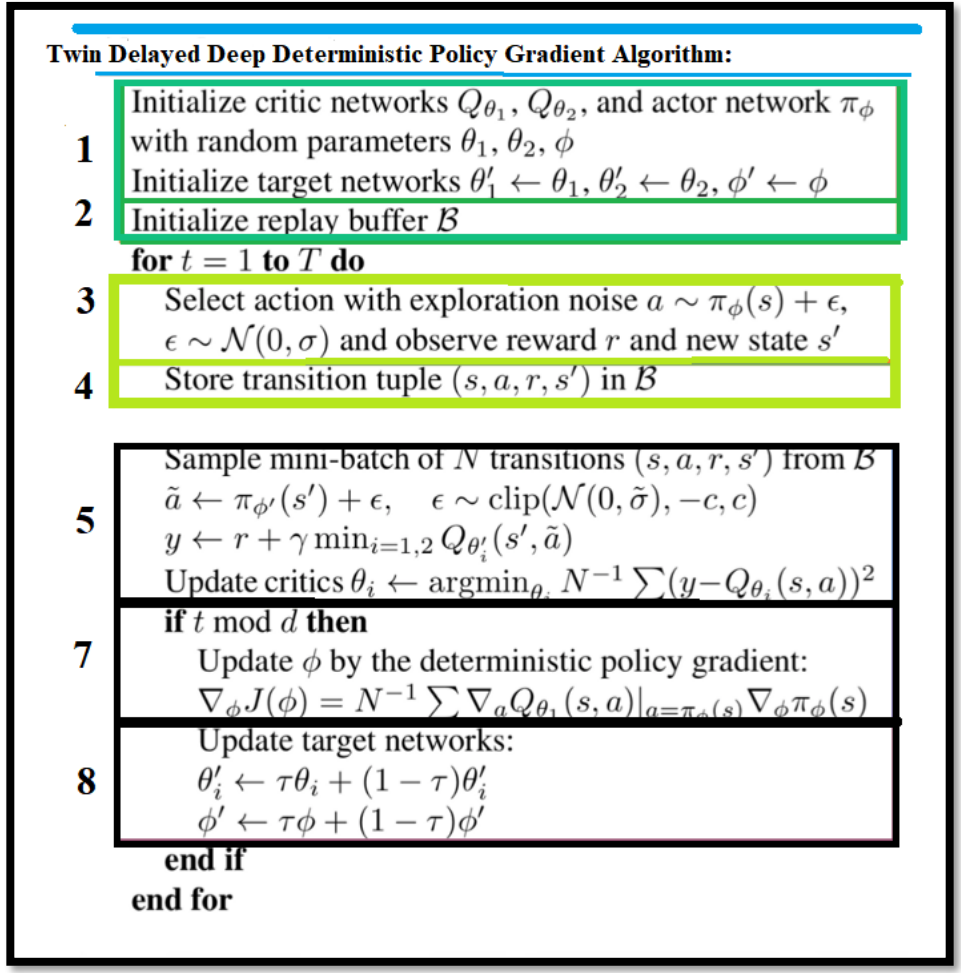

**Fig. 3.** TD3 Algorithm pseudocode

*2.2. TD3 application in closed-loop vibration control for a semi-active suspension system*

The main inputs for vehicle dynamics are road disturbance profile and damping force produced by the MR device. The outputs are body (sprung mass) acceleration and suspension working space (SWS). The RL-Agent inputs for implementing a controller for a one-quarter suspension system are body acceleration, denoted by $q$, and a reward function, which is as follows:

$$r = \begin{cases} 0 & if \ q = q_{goal} = 0 \\ -kq^2 & if \ q \neq 0 \end{cases} \tag{29}$$

Where *k* is a hyperparameter that specifies the intensity coefficient for agent punishment, the agent punishments experienced in the replay buffer are also exerted in the RL-agent. The damper's input voltage must be applied to the coil via the RL-agent output (action). TD3 analyzes its performance by measuring body acceleration and fine-tuning its action to road profile disturbances. The surrounding environment consists of a vehicle suspension system equipped with an MR-damper and a road profile. The agent is a neural network that constructs the controller part. The hyperparameters listed in Table 3 pertain to TD3 agents used in closed-loop suspension control. Figures 5 and 6 detail the neural networks used in the TD3 algorithm.

The hidden size is *400* and *300* neurons in each layer for the actor and critic network, respectively. The optimization method is Adam for both actor and critic networks. The discount factor $\gamma$ is *0.8*. The input voltage varies from *0 to 3 V,* and the damping force varies from *-1.5 to 1.5 KN*.

**Table 3.** Neural Networks Parameters in TD3

| Network | Learning Rate | Optimizer | Delay for update |
|---|---|---|---|
| Actor | 0.002 | Adam | 2 |
| Critic | 0.002 | Adam | 1 |
| Actor Target | 0.006 | - | 2 |
| Critic Target | 0.006 | - | 2 |

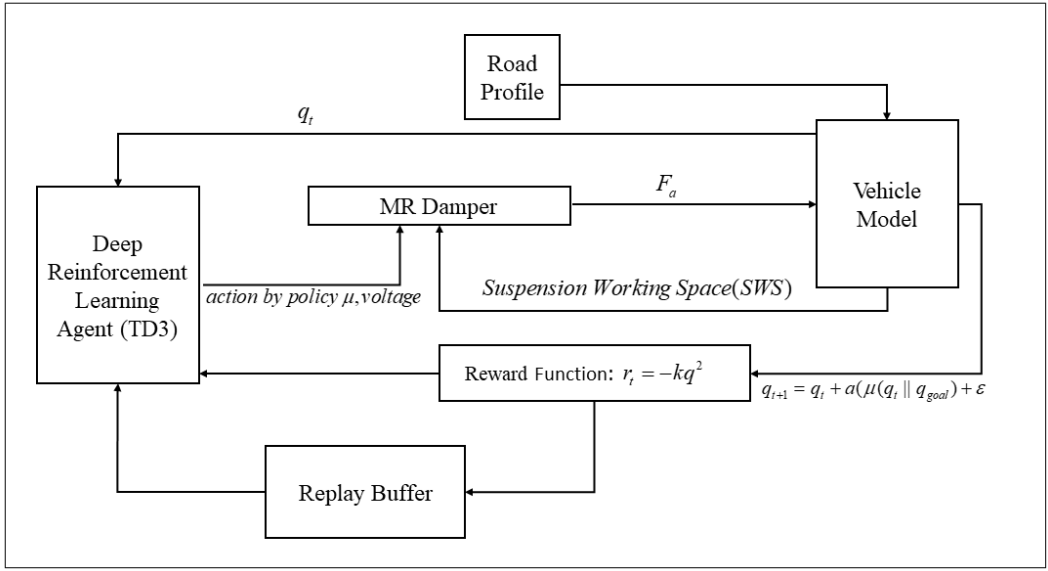

**Fig. 4.** Closed-loop semi-active suspension system block diagram with DRL Agent

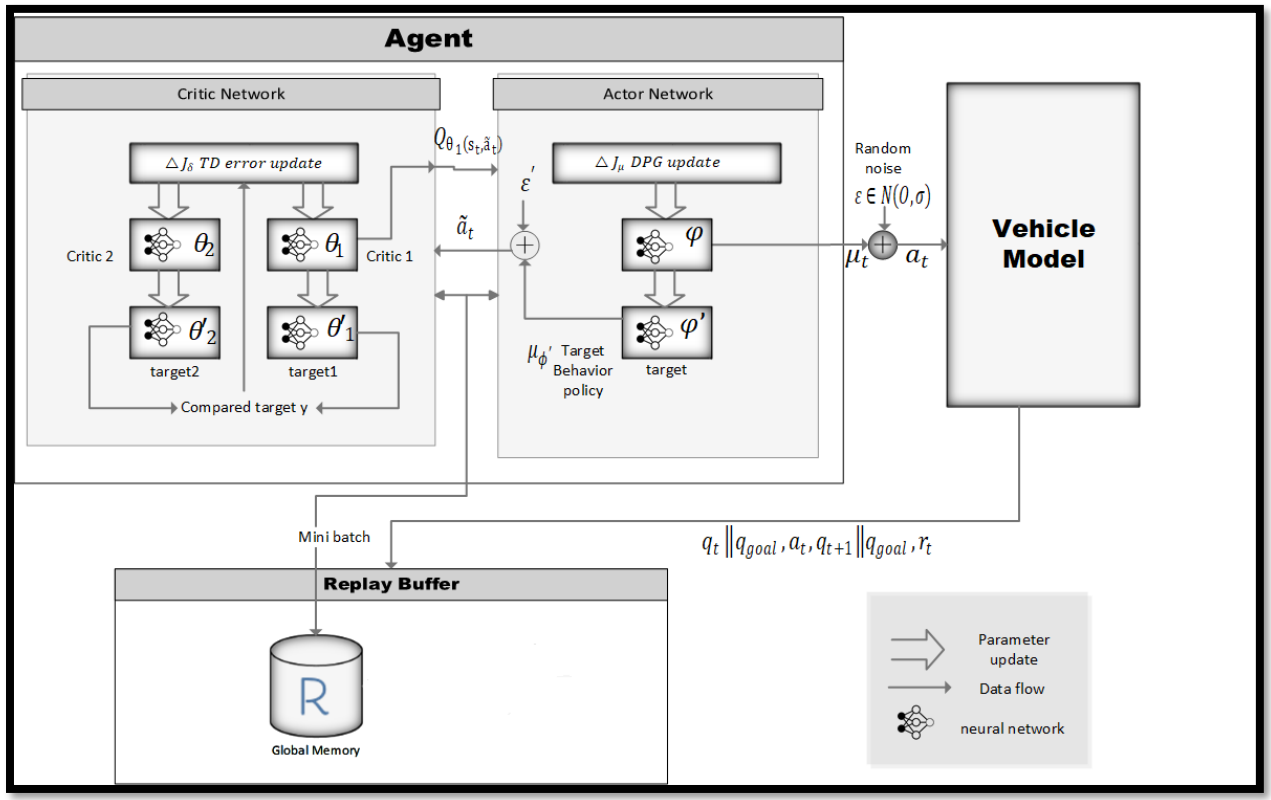

**Fig. 5**. Actor-critic architecture in TD3 Agent

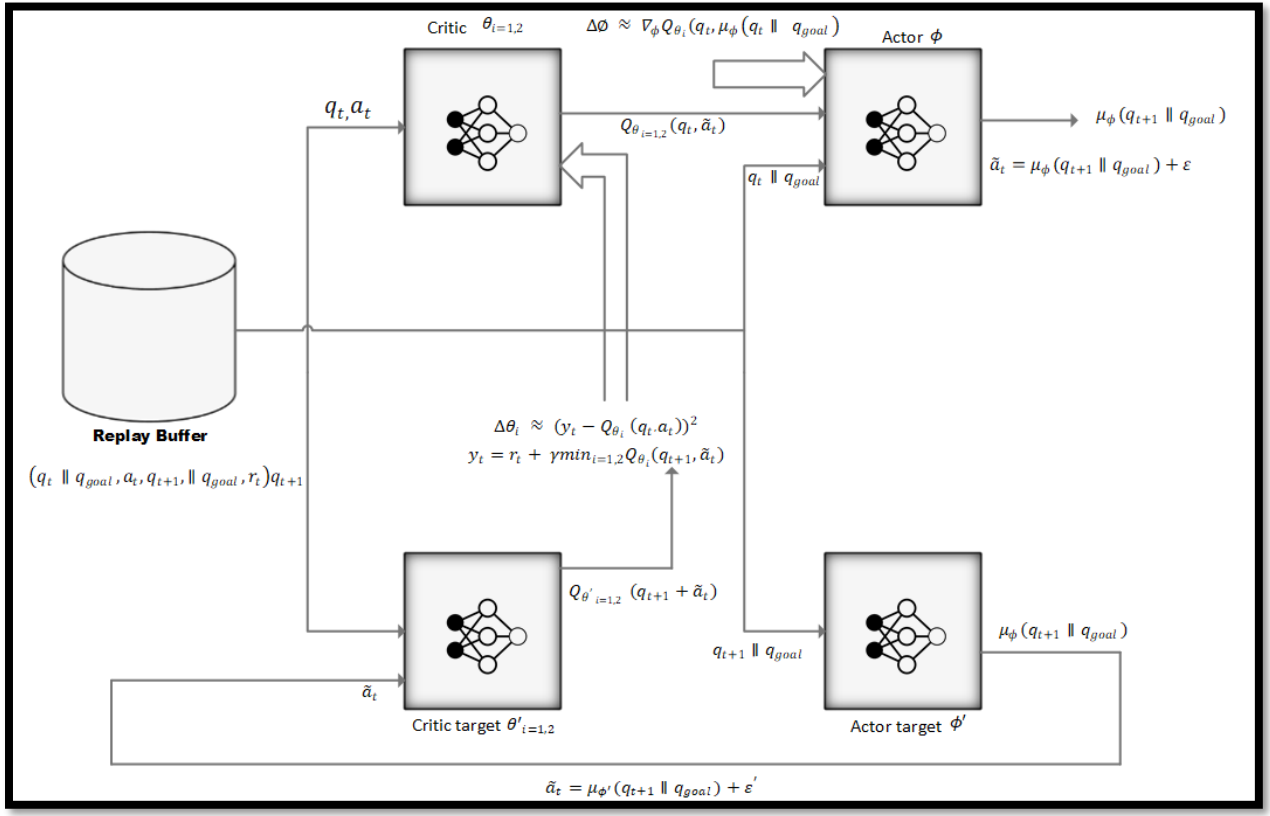

**Fig. 6.** The details of actor-critic networks in semi-active suspension control

## 3. Results and Discussion

Vertical body acceleration (BA), Dynamic Tire Load (DTL), and Suspension Working space are three primary performance criteria used in the suspension design of a vehicle to improve ride comfort and road holding, preventing the suspension system from bottoming out excessively. To achieve the objectives mentioned above, we should reduce BA or SWS to improve ride comfort and DTL to improve road holding and minimize suspension space distance. The BA has been chosen as the objective function in this article.

Two types of controllers for semi-active suspension are investigated in this section: the RL-based TD3 and the PID controller; their gains are tuned using the Particle Swarm Optimization (PSO) algorithm from [17], as well as an uncontrolled system (MR-Passive) with no applied voltage to the damper's coil.

A particular type of bump-road excitation was chosen due to its similarity to actual road profiles. This scenario is a common and standard scenario based on ISO 2631. Furthermore, we believe that the Deep RL algorithm can effectively handle various road profiles and scenarios because it can learn continuously and online. The selected bump-road profile also demonstrates the characteristics of transient response. The road-bump profile has been proposed in [29] as:

$$Xr = \begin{cases} a\{1 - \cos(\omega_r(t - 0.5))\}, for\, 0.5 \leq t \leq 0.5 + \frac{d_b}{V_c} \\ 0, otherwise \end{cases} \quad (30)$$

Where $a$ denotes half of the bump amplitude, $\omega_r = 2\pi \frac{V_c}{d_b}$, $d_b$ is the bump width, and $V_c$ denotes the vehicle velocity. In this research, $a = 0.035m, d_b = 0.8m, V_c = 0.856\frac{m}{s}$, derived from [29].

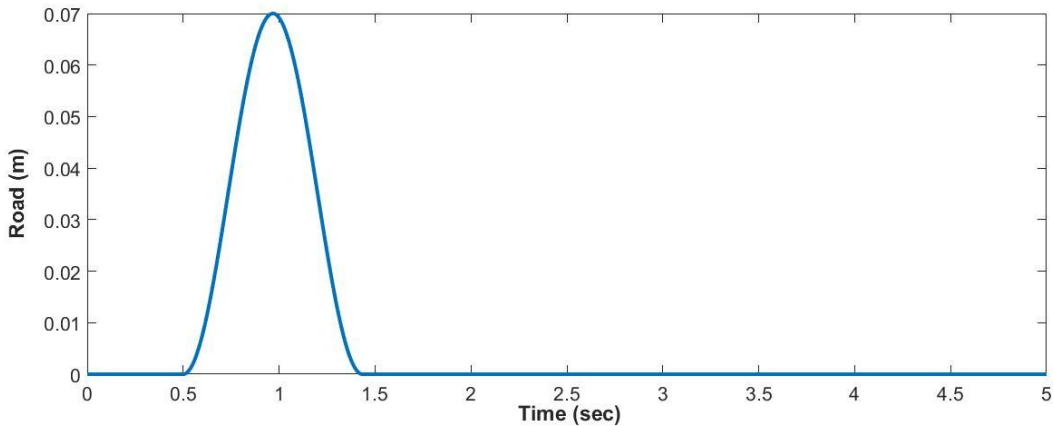

**Fig. 7**. Road-bump profile

Table 4 compares the results from the TD3 controller to an uncontrolled system (applied zero voltage), and Table 5 compares the results from the PSO PID controller to the TD3.

The results demonstrate unequivocally that the DRL-based controller (TD3) algorithm outperforms PSO-tuned PID. TD3 is excellent at dissipating vibrations caused by bump excitation. Additionally, it reduces settling time and enhances road holding and ride comfort.

DRL TD3 significantly decreases body acceleration, suspension displacement, and dynamic tire load in comparison to an uncontrolled suspension system by 35.8%, 68.5%, and 33.6%, respectively. Simultaneously, PSO-PID reduces those criteria by 32.2%, 50%, and 12.4%, respectively, compared to MR passive (uncontrolled suspension).

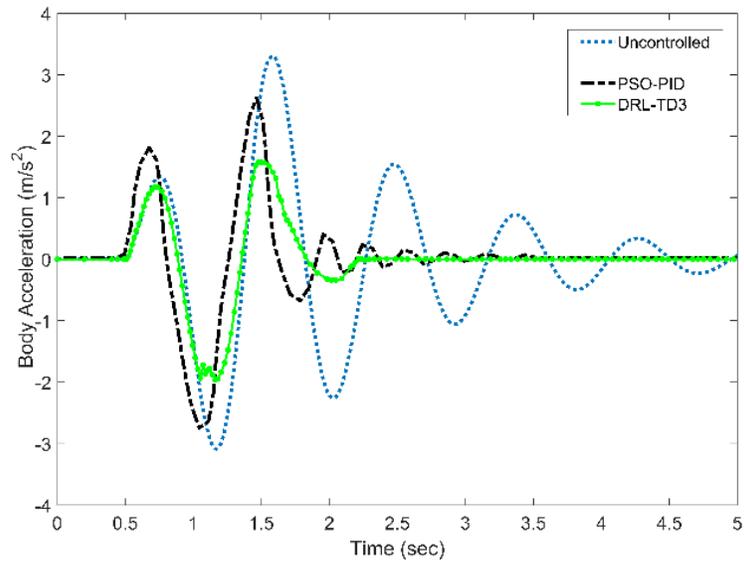

**Fig. 7.** Body Acceleration (BA)

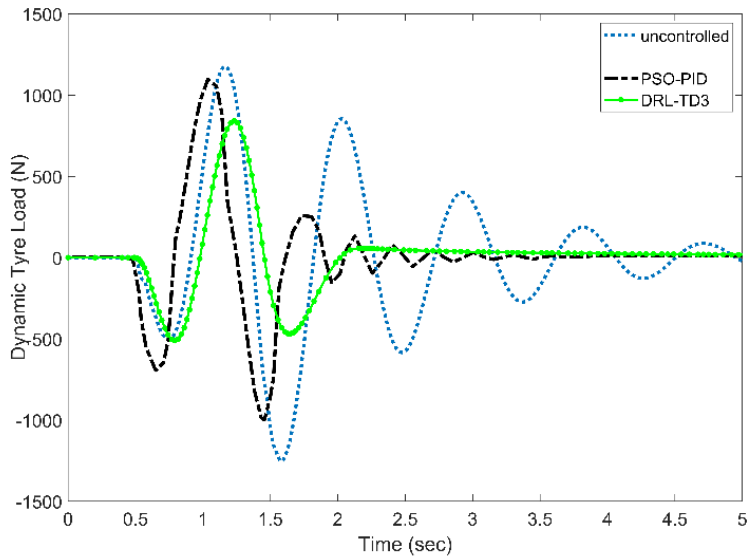

**Fig. 8.** Dynamic Tire Load (DTL)

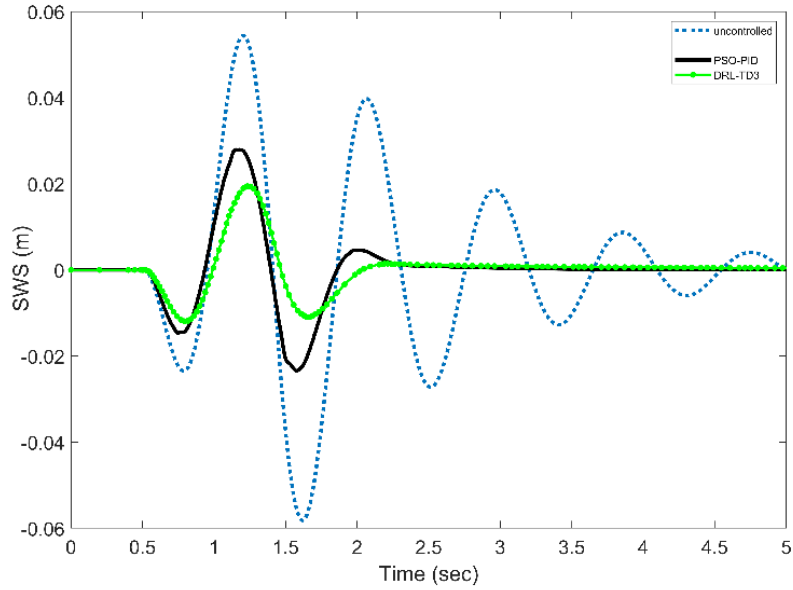

**Fig. 9.** Suspension Working Space (SWS)

**Table 4.** RMS Values for the suspension system criteria with TD3 algorithm and Uncontrolled Suspension System

| RMS-values | Body Acceleration($m/s^2$) | Suspension Working Space($m$) | Dynamic Tire Load ($N$) |
|---|---|---|---|
| *DRL-TD3* | 0.97 | 0.0084 | 381.32 |
| *Uncontrolled* | 1.5179 | 0.0267 | 537.7 |

**Table 5.** Comparison of DRL-TD3 with PSO-PID

| Controller Type | Body Acceleration($m/s^2$) | Suspension Working Space($m$) | Dynamic Tire Load ($N$) |
|---|---|---|---|
| *DRL-TD3* | 35.8 % | 68.53 % | 33.6% |
| *PSO-PID* | 22.4% | 46.1% | 11.9% |
| *Improvement* | 13.4% | 22.43% | 21.7% |

## 4. Conclusion

In this paper, we proposed the use of a deep reinforcement learning (DRL) algorithm to mitigate vibrations in a semi-active suspension system equipped with an MR damper. Our results demonstrated that the DRL-based TD3 algorithm outperformed the PSO-tuned PID controller in terms of three major criteria - BA, SWS, and DTL. Moreover, we introduced permanent online learning for semi-active suspension systems, which can adapt to changing road conditions in real-time, making the system more robust and effective.

Our study highlights the potential of using DRL algorithms for controlling semi-active suspension systems. The results of this study could be applied to various applications such as different type of vehicles, where vibration control is crucial for passenger comfort and safety. We believe that this area of research has significant potential for further exploration, and our findings could be extended to other fields related to control, optimization, and machine learning.

One promising avenue for future research is the incorporation of transformers in the context of reinforcement learning, which has shown promising results in other fields. Transformers can capture long-term dependencies and temporal patterns in the data, enabling better predictions and control actions. Additionally, the use of computer vision for analyzing road conditions could provide valuable information for developing more robust control strategies. Computer vision techniques can extract features from the images or videos of the road surface, which can help in predicting the road condition and adjusting the control parameters accordingly.

In conclusion, our study demonstrates the effectiveness of the DRL-based TD3 algorithm for mitigating vibrations in semi-active suspension systems equipped with MR dampers. We hope that our findings contribute to the development of more advanced control strategies for semi-active suspension systems and inspire further research in this area. The potential of using DRL algorithms in various applications and the incorporation of transformers and computer vision for future research could lead to breakthroughs in the field of control and optimization.

**Declaration of competing interest**

The authors affirm that they do not have any known competing financial interests or personal relationships that could have influenced the work presented in this paper.